\documentclass[aps,pra,showpacs,twocolumn]{revtex4}

\usepackage{graphicx}

\begin{document}

\title{Complete characterization of post-selected quantum statistics 
\\ using weak measurement tomography}

\author{Holger F. Hofmann}
\email{hofmann@hiroshima-u.ac.jp}
\affiliation{
Graduate School of Advanced Sciences of Matter, Hiroshima University,
Kagamiyama 1-3-1, Higashi Hiroshima 739-8530, Japan}

\begin{abstract}
The reconstruction of quantum states from a sufficient set of experimental data can be achieved with arbitrarily weak measurement interactions. Since such weak measurements have negligible back-action, the quantum state reconstruction is also valid for the post-selected sub-ensembles usually considered in weak measurement paradoxes. It is shown that post-selection can then be identified with a statistical decomposition of the initial density matrix into transient density matrices conditioned by the anticipated measurement outcomes. This result indicates that it is
possible to ascribe the properties determined by the final measurement outcome to each individual quantum system before the measurement has taken place. The ``collapse'' of 
the pure state wavefunction in a measurement can then be understood in terms of the classical 
``collapse'' of a probability distribution as new information becomes available. 
\end{abstract}


\maketitle

\section{Introduction}

As our ability to control individual quantum systems increases, the seemingly paradoxical aspects of quantum measurement theory take on a more practical relevance. One recent example is the renewed interest in the quantum statistics of post-selected weak measurements, which seem to suggest the presence of negative probabilities as the source of quantum paradoxes 
\cite{Aha02,Res04a,Mir07,Jor06,Wil08,Gog09,Lun09,Yok09}. These developments in the field of weak 
measurement may be especially significant in the context of new experimental possibilities pioneered by quantum information related research \cite{Pry05,Ral06,Men08}. However, there seems to be a certain mismatch between the conventional approach to weak measurements, which is based on the notion of measurement interaction dynamics mediated by operator observables \cite{Aha88,Res04b,Joh04,Jos07,Shp08}, and the more general approach to measurements based on the measurement operators widely used in quantum information \cite{NieCh}. In particular, the conventional analysis seems to overemphasize the exotic and surprising aspects of the weak measurements, while the operator based approach shows more clearly how weak measurements fit into the general framework of quantum physics \cite{Ore05}. As experimental weak measurements become more and more established, it may therefore be time to shift the focus away from the oddities of specific cases, towards a complete and consistent formulation of post-selection effects in terms of their experimentally observable properties.

From the experimental side, quantum states and processes can be characterized by measuring their complete statistical properties, a procedure known as quantum tomography \cite{Leo95,Whi99}. A significant merit of quantum tomography is that it establishes an operational approach to quantum states, that is, it defines the quantum state in terms of the experimentally accessible data. By applying quantum tomography to generalized weak measurements, it is possible to extend this operational definition to post-selected sub-ensembles of a quantum state.  
In the following, the general theory of quantum tomography with measurements of variable strength is formulated. It is shown that, in the limit of weak measurements, post-selection partitions the initial density matrix into sub-ensembles described by non-positive transient density matrices. This result divides the ``collapse'' of the wavefunction into two distinct parts, one associated with the selection of the appropriate sub-ensemble, and the other related to the actual back-action of the measurement dynamics. Although the measurement back-action is 
necessary to cover up the negative eigenvalues of the transient states, weak measurement tomography suggests that the sub-ensemble partition has a physical meaning even before the measurement interaction takes place. 

\section{Quantum state tomography with weak measurements}

The starting point for the following derivation of a complete and consistent theory of weak measurement tomography is the representation of general quantum measurements by a set of operators $\{ \hat{M}_m \}$ acting only on the Hilbert space of the system \cite{NieCh}. These operators summarize the relevant effects associated with a measurement outcome $m$, separating the essential properties of a quantum measurement from the technical problem of 
its implementation by a specific combination of system-meter interaction, meter preparation, 
and meter read-out.  In the case of Hilbert space vectors representing pure states, the application of a measurement operator $\hat{M}_m$ to a state vector changes both the length and the direction of that vector. The new direction of the state vector then represents the output state after the measurement, while the squared length represents the probability $p(m)$ of obtaining the measurement outcome $m$. If the quantum state is expressed in terms of the density matrix $\hat{\rho}_i$, the probability $p(m)$ is given by a product trace,
\begin{equation}
\label{eq:POVM}
p(m)= \mbox{Tr}\left\{\hat{M}_m \; \hat{\rho}_i \hat{M}_m^\dagger \right\}=\mbox{Tr}\left\{\hat{M}_m^\dagger \hat{M}_m \; \hat{\rho}_i \right\}.
\end{equation}
The set of squared operators $\{\hat{M}_m^\dagger \hat{M}_m\}$ is the positive operator-valued measure (POVM) of the measurement probabilities. Since all probabilities must add up to $1$, the POVM fulfills the completeness relation
\begin{equation}
\label{eq:complete}
\sum_m \hat{M}_m^\dagger \hat{M}_m = \hat{1}.
\end{equation}
The POVM formalism describes the general relation between experimental data and the quantum state. Specifically, eq.(\ref{eq:POVM}) shows that the measurement probabilities $p(m)$ are linear combinations of the density matrix elements. If the set of relations given by eq.(\ref{eq:POVM}) is invertible, the complete density matrix can be reconstructed from the available set of measurement probabilities. In a $d$-dimensional Hilbert space, quantum tomography can thus be performed using any combination of $d^2$ linearly independent measurement operators $\hat{M}_m^\dagger \hat{M}_m$.

Whether a POVM is invertible and therefore suitable for quantum tomography does
not depend on the precision of the measurement. It is therefore possible to 
reconstruct the quantum state from arbitrarily weak measurements. To illustrate
this point, it is useful to formulate the inversion procedure for POVMs with variable strength $\epsilon$. 
The strength or weakness of a measurement can be quantified directly by the closeness of the measurement operators to multiples of the identity operator $\hat{1}$. A convenient way of representing a variably measurement strength $\epsilon$ in the formalism is
\[
\hat{M}_m^\dagger \hat{M}_m = w_m (\hat{1}+\epsilon \hat{S}_m)
\]
\vspace*{-0.5cm}
\begin{equation}
\label{eq:Sform}
\mbox{with} \hspace{0.5cm}
\sum_m w_m = \frac{1}{1+\epsilon} \hspace{0.3cm}
\mbox{and} \hspace{0.3cm} 
\sum_m w_m \hat{S}_m = \frac{\hat{1}}{1+\epsilon}.
\end{equation}
The measurement probabilities in eq. (\ref{eq:POVM}) can then be expressed in terms of the expectation values of a set of self-adjoined operators $\hat{S}_m$
that is independent of the measurement strength $\epsilon$.
For quantum state reconstruction, a set of $d^2$ linearly independent operators $\{\hat{S}_m\}$ defines an operator expansion of the density matrix in terms of the set of reciprocal operators $\{\hat{\Lambda}_m\}$ with
$\mbox{Tr}\{\hat{S}_n \hat{\Lambda}_m\}=\delta_{n,m}$. The density matrix is then given by
\begin{equation}
\label{eq:tomography}
\hat{\rho}_i=\sum_m \mbox{Tr}\{\hat{S}_m \hat{\rho}_i\} \hat{\Lambda}_m,
\end{equation}
where the coefficients of the expansion are related to the measurement probabilities $p(m)$ by
\begin{equation}
\label{eq:reconstruct}
\mbox{Tr}\{\hat{S}_m \hat{\rho}_i\}= \frac{1}{\epsilon w_m}
\left( p(m)-w_m \right).
\end{equation}
Eqs. (\ref{eq:tomography}) and (\ref{eq:reconstruct}) express the density matrix in terms of the experimentally obtained measurement probabilities $p(m)$. Thus, quantum tomography can provide a definition of the density matrix that is based only on empirical properties of 
the system.

\section{Quantum states of post-selected ensembles}

Conventional quantum tomography is a one-way readout process in which the quantum state is discarded after the measurement. However, the measurement operators $\hat{M}_m$ also provide a description of the quantum state after the measurement. It is therefore possible to consider the effects of a subsequent
measurement with outcomes $f$, described by another POVM $\{\hat{\Pi}_f \}$. 
The joint probability of obtaining first $m$ and then $f$ in the measurements is given by
\begin{equation}
\label{eq:pjoint}
p(m,f|i) = \mbox{Tr}\left\{\hat{\Pi}_f  \hat{M}_m \hat{\rho}_i \hat{M}_m^\dagger  \right\}.
\end{equation}
Since the measurement operators $\hat{M}_m^\dagger$ and $\hat{M}_m$ are not directly multiplied, the probability $p(f|i)$ found by summing over all $m$ is different from the product trace of $\hat{\rho}_i$ and $\hat{\Pi}_f$. This change
in the probability of obtaining $f$ from the initial state $\hat{\rho}_i$ is caused by the measurement back-action associated with the measurement of $m$. 
It is therefore impossible to know whether the final result $f$ was caused by the initial state of the system or by interaction effects related to the measurement outcome $m$. However, the theory of weak measurements shows how this problem can be circumvented: for very small measurement strengths $\epsilon$, the effects of the quantum state on the measurement probabilities is linear in $\epsilon$ while the measurement back-action is quadratic in $\epsilon$. It is therefore possible to realize quantum state tomography with negligible back-action.

In the limit of weak measurements ($\epsilon \ll 1$), the measurement operators $\hat{M}_m$ are approximately given by the linearized square root of the POVM,
\begin{equation}
\label{eq:weakM}
\hat{M}_m \approx \sqrt{w_m}\left(\hat{1}+\frac{\epsilon}{2} \hat{S}_m\right).
\end{equation}
In the joint probability $p(m,f|i)$ given by eq.(\ref{eq:pjoint}), the terms 
linear in $\epsilon$ are obtained by applying $\hat{S}_m$ either from the 
right or from the left. If quadratic terms are neglected, this is equivalent 
to applying the POVM from the right or from the left. Therefore, the joint
probability $p(m,f|i)$ can be approximately expressed by the 
weak measurement POVM,
\begin{equation}
\label{eq:wmapprox} 
p(m,f|i) \approx \mbox{Tr}\left\{
\hat{\Pi}_f \; \frac{1}{2}(\hat{\rho}_i \hat{M}_m^\dagger \hat{M}_m 
+ \hat{M}_m^\dagger \hat{M}_m \hat{\rho}_i)\right\}.
\end{equation}
Since the POVM fulfills the completeness relation given by eq.
(\ref{eq:complete}), the final measurement probabilities $p(f|i)$ are 
not changed by the measurement of $m$ and 
\begin{equation}
\label{eq:pfi}
p(f|i)=\sum_m p(m,f|i)=\mbox{Tr}\{\hat{\Pi}_f \hat{\rho}_i \}.
\end{equation}
Hence the measurement back-action is negligible and the measurement results
$m$ merely identify the state conditioned by both initial preparation and final measurement. 
The conditional probability can then be written in terms of a transient density matrix 
$\hat{R}_{if}$, such that
\begin{equation}
\label{eq:mfiR}
p(m|i,f) = \frac{p(m,f|i)}{p(f|i)}=
\mbox{Tr}\left\{\hat{M}_m^\dagger \hat{M}_m \hat{R}_{if} \right\}.
\end{equation}
According to eqs.(\ref{eq:wmapprox}) and (\ref{eq:pfi}), this transient 
density matrix is given by
\begin{equation}
\hat{R}_{if} = \frac{1}{2 \mbox{Tr}\{\hat{\rho}_i\hat{\Pi}_f\}}
\left(\hat{\rho}_i \hat{\Pi}_f + \hat{\Pi}_f \hat{\rho}_i \right).
\end{equation}
Since quantum tomography provides an unambiguous definition of the density matrix
in terms of the measurement statistics of $m$ for any well-defined ensemble of 
quantum systems, the consistency of quantum measurement theory requires that the 
self-adjoint operator $\hat{R}_{if}$ is the proper statistical representation of 
the sub-ensemble defined by the conditions $i$ and $f$. 

\section{Anticipatory decomposition}

The operator $\hat{R}_{if}$ correctly predicts the outcomes of all (real) weak values that
can be obtained between $i$ and $f$. In this sense, it is similar to the
two-state vector formalism of weak measurements \cite{Aha01} and its extension to
mixed states \cite{Shi08}. However, the derivation from tomography ensures that
$\hat{R}_{if}$ is a self-adjoined operator with real eigenvalues and a trace of one. 
It thus corresponds to a conventional density matrix, except for the possibility of 
negative eigenvalues. 
Moreover, the definition of $\hat{R}_{if}$ applies equally well to projective measurements
of pure states and to noisy measurements of mixed states. The present analysis therefore bridges the gap between
the classical limit and the extreme quantum limit, revealing similarities 
of quantum and classical statistics that tend to be obscured by the
state vector formalism. Specifically, the transient density matrix $\hat{R}_{if}$
is the quantum mechanical representation of conditional probabilities 
$p(m|i,f)$, just as the density matrix $\hat{\rho}_i$ is the quantum mechanical 
representation of the probabilities $p(m|i)$. The relation between the total density 
matrix $\hat{\rho}_i$ and the set of transient density matrices conditioned by the final
measurement outcomes $f$ is therefore given by the weighted sum over all possible outcomes $f$,
\begin{equation}
\label{eq:decomp}
\hat{\rho}_i = \sum_f p(f|i) \; \hat{R}_{if}.
\end{equation}
This decomposition of the density matrix explains why the weak measurement 
statistics can be measured before the final measurement $f$ has taken place:
the measurement of $f$ simply identifies a sub-ensemble $\hat{R}_{if}$ that 
is already included in the total ensemble $\hat{\rho}_i$. It is therefore
possible to decompose $\hat{\rho}_i$ into $\hat{R}_{if}$ in {\it anticipation} 
of the measurement of $f$. Such an anticipatory decomposition indicates that the
measurement outcome $f$ can already be ascribed to the quantum systems before 
the measurement has taken place. 

\section{Wavefunction collapse without measurement back-action}

In classical physics, the statistical rules for anticipatory decompositions
are straightforward, since the state $\hat{\rho}$ is just a conventional 
probability distribution over all microscopic configurations of the system.
In the quantum case, it is usually assumed that the ``collapse'' of a pure 
state caused by a projective measurement is fundamentally different from
such a probability update, because it includes the effects of decoherence.
However, weak measurement tomography shows that the ``collapse'' can be
divided into two parts, one associated with a classical sub-ensemble selection
(that is, a Bayesian probability update), and the second one associated with 
the actual physical interaction that results 
in the decoherence. It is therefore possible to identify the set of quantum systems
that produce a specific measurement result $f$ with a uniquely defined
sub-ensemble of the total density matrix $\hat{\rho}_i$. 

To understand the significance of this result, it may be useful to
reconsider some of the paradoxes of quantum mechanics in the light of
these experimentally accessible facts. For instance, it is now possible
to resolve the problem of double slit interference by assigning both
an interference pattern and a slit to each individual particle. 
Specifically, the initial pure state superposition 
$\mid i \rangle=(\mid 1 \rangle
+ \mid 2 \rangle)/\sqrt{2}$ of a particle passing through slit 1
and a particle passing through slit 2 can be decomposed into 
\begin{eqnarray}
\label{eq:dslit}
\hat{R}_{i1} &=& \mid 1 \rangle \langle 1 \mid +
\frac{1}{2}\left(\mid 1 \rangle \langle 2 \mid
+ \mid 2 \rangle \langle 1 \mid \right)
\nonumber \\
\hat{R}_{i2} &=& \mid 2 \rangle \langle 2 \mid +
\frac{1}{2}\left(\mid 1 \rangle \langle 2 \mid
+ \mid 2 \rangle \langle 1 \mid \right)
\end{eqnarray}
in anticipation of a which-path measurement. Weak measurements performed
between the preparation of the superposition $\mid i \rangle$ and 
the final which-path measurements show that the path information
co-exists with the interference pattern of the superposition. Therefore,
the double slit interference pattern can be obtained from weak measurements 
even if the particle has passed through only one of the slits. 
Eq. (\ref{eq:dslit}) thus shows that coherence between two alternatives does 
not imply that both alternatives are simultaneously realized. 

Essentially, weak measurement tomography is an analysis of quantum statistics
that reveals additional details about where quantum information is located
before it is measured. In particular, it fills a gap left
by the measurement postulate by showing that the measurement result $f$
corresponds to properties of the system before the measurement, and is not
just randomly generated by a role of the dice at the time of measurement. 
Thus, we can solve the paradox of Schr\"odinger's cat by experimentally
confirming that the cat was already dead before somebody opened the
box to look. Likewise, we can conclude that the non-local collapse of
entangled states merely corresponds to a Bayesian probability update in the
remote system, providing a classical analogy that can explain non-classical
properties of quantum mechanics such as the impossibility of non-local 
signaling and the transfer of quantum information by classical channels 
in quantum teleportation \cite{Hof00,Hof02}.

\section{Experimentally observable negative probabilities}

As explained above, weak measurement tomography identifies the quantum 
statistics of systems with well-defined initial and final properties. 
This means that the available information about each system can be more precise than the
uncertainty limit allows. Such super-certain information finds its quantum 
statistical expression in the non-positive density matrices $\hat{R}_{if}$ 
that summarize the results of all possible weak measurements between $i$ and $f$.
The negative eigenvalues of $\hat{R}_{if}$ represent weak values that exceed the 
eigenvalue bounds of positive density matrices,
providing a consistent framework for the resolution of quantum paradoxes by
weak measurements \cite{Jor06,Wil08,Gog09,Lun09,Yok09}. 

Quantitative descriptions of quantum paradoxes such as Bell's inequalities \cite{Bel64}, 
Leggett-Garg inequalities \cite{Jor06,Wil08,Gog09} or contextuality inequalities 
\cite{Cab08} are usually
formulated in terms of precise measurements performed on different representatives 
of the same state. To connect this conventional formulation with weak measurements, it
is necessary to show that the joint probabilities determined in weak measurements 
provide a unique and measurement independent definition of joint probabilities in
quantum systems. 
In particular, the joint probabilities should not depend on the order 
in which the result is obtained \cite{Rau09}. For a pair of strong measurements 
$\{\hat{\Pi}_f\}$ 
and $\{\hat{\Phi}_g\}$, the joint probability from a weak measurement of $g$ followed by a 
strong measurement of $f$ should therefore be equal to the joint probability of a
weak measurement of $f$ followed by a strong measurement of $g$,
\begin{eqnarray}
\label{eq:fgcond}
p(f,g|i) &=& p(f|i) \,\mbox{Tr}\{\hat{\Phi}_g \hat{R}_{if}\} 
\nonumber \\      
         &=& p(g|i) \,\mbox{Tr}\{\hat{\Pi}_f \hat{R}_{ig}\}. 
\end{eqnarray}
Weak measurement tomography can confirm that these two results are indeed equal.
This means that the transient density matrices $\hat{R}_{if}$ and $\hat{R}_{ig}$
describe the same statistical correlations between the measurement of $f$ and
the measurement of $g$. We can therefore conclude that weak measurement tomography
provides a consistent definition of joint probabilities for measurements that cannot
be performed jointly. In terms of the POVMs, this joint probability reads
\begin{equation}
\label{eq:fgjoint}
p(f,g|i)= \mbox{Tr}\left\{\frac{1}{2}\left(\hat{\Phi}_g \hat{\Pi}_f
+ \hat{\Pi}_f \hat{\Phi}_g \right) \; \hat{\rho}_i
\right\}.
\end{equation}
It should be emphasized that each of these joint probabilities can be measured
directly in an appropriate weak measurement. Eq.(\ref{eq:fgjoint}) is therefore
the only definition of joint probability with empirical validity. Interestingly, this
means that quasi-probabilities such as the Wigner function should not be interpreted in
terms of joint probabilities, since their mathematical construction does not 
refer to a valid pair of measurements \cite{note}.
On the other hand, the negative probabilities predicted by eq.(\ref{eq:fgjoint}) are a 
natural consequence of quantum statistics, required by the consistency of weak and 
strong results. Thus, the validity of the results is not a matter of interpretation,
even though the precise meaning of negative joint probabilities for outcomes that never 
occur jointly may be difficult to understand. 

\section{Implications of uncertainty for individual quantum systems}

Obviously, negative probabilities cannot be interpreted as relative frequencies of
actual measurements. Nevertheless, they can be derived from the relative
frequencies of weak measurements that consistently give the same results as the 
corresponding strong measurements. To reconcile the strangeness of negative
probabilities with the empirical sense of reality justified by reproducible
measurement results, it is important to remember that the validity of an 
anticipatory decomposition depends on the performance 
of the actual measurement. If an alternative measurement is performed instead, 
the sub-ensembles need to be
``re-shuffled'' before the correct decomposition is applied. For practical
purposes, contradictions are avoided because the reality of an individual 
system is determined by only one of the possible measurements. The resolution
of quantum paradoxes by negative probabilities is therefore based on the difference 
between the total statistical ensemble and its individual representatives. The 
reality of the representative is given by precise values of $i$ and $f$ (or $i$ and $g$),
while the ensemble properties are described by the density matrix. In the case
of the sub-ensemble defined by $i$ and $f$, this density matrix is given by
$\hat{R}_{if}$, which may have negative eigenvalues since it can only be 
observed in weak measurements. Eqs. (\ref{eq:fgjoint}) show that
the joint negative probabilities predicted by weak measurement tomography of $i$ and $f$ 
are consistent with the joint negative probabilities obtained from $i$ and $g$. The result 
can therefore be summarized in terms of a uniquely defined joint probability 
(\ref{eq:fgjoint}) that expresses the relation between strong measurements of $f$ and 
of $g$ directly in terms of their POVM operators. However, the negative eigenvalues of
this operator clearly indicate that individual realities must be restricted to the
actual measurement outcomes caused by the respective system. Weak measurement tomography
thus decides quantum paradoxes in favor of locality, causality, and non-contextuality,
but against the notion of a non-empirical realism that attempts to provide a description of
individual quantum objects beyond the uncertainty limited reality of its individual 
effects. 

\section{Conclusions}

In conclusion, weak measurement tomography reveals a striking consistency
of the quantum measurement formalism with classical statistics by defining an
unambiguous partition of the total ensemble described by $\hat{\rho}_i$ into 
well-defined sub-ensembles $\hat{R}_{if}$ representing the future measurement outcomes
$f$. The reconstruction of quantum states by weak measurements thus provides empirical
evidence that the selection of a measurement outcome $f$ does not eliminate the 
coherences between $f$ and other outcomes. 
In particular, weak measurements can show that particles moving only through slit 1
of a double slit experiment carry the complete interference pattern of the initial
state with them until the physical interaction of the final measurement randomizes
the phase relation. Empirical evidence thus favors a statistical interpretation of
quantum mechanics that assigns reality to individual measurement outcomes even before
the measurement is performed. Quantum paradoxes can then be explained in terms of
the negative probabilities described by the non-positive density matrices $\hat{R}_{if}$.
As was shown above, these transient density matrices uniquely define the
joint probabilities between the measurement results $f$ and the possible outcomes 
of other measurements $g$. The negative values of such joint probabilities demonstrate 
that no consistent simultaneous assignment of both $g$ and $f$ is possible. 
Weak measurements can thus provide experimental proof that quantum paradoxes arise
from the fallacious imposition of non-empirical realities beyond the uncertainty 
limits restricting the observable effects of individual systems.

\section*{Acknowledgment}

I would like to thank Prof. A. Hosoya for pointing out the possibility of 
using weak measurements to perform quantum state tomography on post-selected states. 
Part of this work has been supported by the Grant-in-Aid program of the Japanese 
Society for the Promotion of Science, JSPS.



\begin{thebibliography}{xyz00}


\bibitem{Aha02}
Y. Aharonov, A. Botero, S. Popescu, B. Reznik, and J. Tollaksen, Phys. Lett. A {\bf 301},
130 (2002).

\bibitem{Res04a}
K.J. Resch, J.S. Lundeen, and A. M. Steinberg, Phys. Lett. A {\bf 324}, 125 (2004).

\bibitem{Mir07} 
R. Mir, J. S. Lundeen, M. W. Mitchell, A. M. Steinberg, J. L. Garretson, and H. M. Wiseman,
New J. Phys. {\bf 9}, 287 (2007).


\bibitem{Jor06}
A. N. Jordan, A. N. Korotkov, and M. B\"uttiker, Phys. Rev. Lett. {\bf 97}, 026805 (2006).

\bibitem{Wil08}
N. S. Williams and A. N. Jordan, 
Phys. Rev. Lett. {\bf 100}, 026804 (2008).

\bibitem{Gog09}
M. E. Goggin, M. P. Almeida, M. Barbieri, B. P. Lanyon, J. L. O'Brien, A. G. White, and
G. J. Pryde, e-print arXiv: 0907.1679v1.


\bibitem{Lun09}
J. S. Lundeen and A. M. Steinberg, Phys. Rev. Lett. {\bf 102}, 020404 (2009).

\bibitem{Yok09}
K. Yokota, T. Yamamoto, M. Koashi, and N. Imoto, New J. Phys. {\bf 11}, 033011 (2009).


\bibitem{Pry05}
G. J. Pryde, J. L. O'Brien, A. G. White, T. C. Ralph, and H. M. Wiseman,
Phys. Rev. Lett. {\bf 94}, 220405 (2005).

\bibitem{Ral06}
T. C. Ralph, S. D. Bartlett, J. L. O'Brien, G. J. Pryde, and H. M. Wiseman,
Phys. Rev. A {\bf 73}, 012113 (2006).


\bibitem{Men08}
D. Menzies and N. Korolkova, 
Phys. Rev. A {\bf 77}, 062105 (2008).

\bibitem{Aha88}
Y. Aharonov, D. Z. Albert, and L. Vaidman, Phys. Rev. Lett. {\bf 60}, 1351 (1988).

\bibitem{Res04b}
K. J. Resch and A. M. Steinberg, Phys. Rev. Lett. {\bf 92}, 130402 (2004).

\bibitem{Joh04}
L. M. Johansen, Phys. Rev. Lett. {\bf 93}, 120402 (2004).

\bibitem{Jos07}
R. Jozsa, Phys. Rev. A {\bf 76}, 044103 (2007).

\bibitem{Shp08}
V. Shpitalnik, Y. Gefen, and A. Romito,
Phys. Rev. Lett. 101, 226802 (2008).

\bibitem{NieCh}
M. A. Nielsen and I. L. Chuang, {\it Quantum Computation and Quantum Information} 
(Cambridge University Press, Cambridge, 2000).


\bibitem{Ore05}
O. Oreshkov and T. A. Brun, Phys. Rev. Lett. {\bf 95}, 110409 (2005).


\bibitem{Leo95}
U. Leonhardt,
Phys. Rev. Lett. {\bf 74}, 4101 (1995).

\bibitem{Whi99}
A. G. White, D. F. V. James, P. H. Eberhard, and P. G. Kwiat,
Phys. Rev. Lett. 83, 3103 (1999).


\bibitem{Aha01}
Y. Aharonov and L. Vaidman, e-print arXiv: quant-ph/0105101v2.

\bibitem{Shi08}
Y. Shikano and A. Hosoya, e-print arXiv: 0812.4502v1.


\bibitem{Hof00}
H. F. Hofmann, T. Ide, T. Kobayashi, and A. Furusawa, Phys. Rev. A {\bf 62},
062304 (2000).

\bibitem{Hof02}
H. F. Hofmann, Phys. Rev. A {\bf 66}, 032317 (2002).


\bibitem{Bel64}
J. S. Bell, Physics {\bf 1}, 195 (1964).


\bibitem{Cab08}
A. Cabello, Phys. Rev. Lett. {\bf 101}, 210401 (2008).


\bibitem{Rau09}
This condition corresponds to the ``reductionism'' required by Bayesian
interpretations of quantum probability, as explained in J. Rau, e-print arXiv: 0710.2119v2.


\bibitem{note}
In particular, weak measurement tomography indicates that the appearance of interferences
at a position between the two slits in the Wigner function of double slit interference
is a mathematical artefact. This explains why these terms can pass through the screen
even where there are no slits. 

\end{thebibliography}
\end{document}